\documentclass[%
reprint,
superscriptaddress,
showpacs,
showpacs,preprintnumbers,
amsmath,amssymb,
aps,
]{revtex4-1}

\usepackage[dvipdfmx]{graphicx}% Include figure files
\usepackage{dcolumn}% Align table columns on decimal point
\usepackage{bm}% bold math
\usepackage{multirow}
\usepackage[mathlines]{lineno}% Enable numbering of text and display math
\usepackage{ulem}
\usepackage{color}
\begin{document}

\preprint{APS/123-QED}

\title{Magneto-transport properties of tellurium under extreme conditions}

\author{Kazuto~Akiba}
\email{akb@okayama-u.ac.jp}
\affiliation{
Graduate School of Natural Science and Technology, Okayama University,
Okayama 700-8530, Japan
}
\affiliation{
The Institute for Solid State Physics, The University of Tokyo,
Chiba 277-8581, Japan
}

\author{Kaya~Kobayashi}
\affiliation{
Research Institute for Interdisciplinary Science, Okayama University,
Okayama 700-8530, Japan
}

\author{Tatsuo C. Kobayashi}
\affiliation{
Graduate School of Natural Science and Technology, Okayama University,
Okayama 700-8530, Japan
}

\author{Ryo~Koezuka}
\affiliation{
The Institute for Solid State Physics, The University of Tokyo,
Chiba 277-8581, Japan
}

\author{Atsushi~Miyake}
\affiliation{
The Institute for Solid State Physics, The University of Tokyo,
Chiba 277-8581, Japan
}

\author{Jun~Gouchi}
\affiliation{
The Institute for Solid State Physics, The University of Tokyo,
Chiba 277-8581, Japan
}

\author{Yoshiya~Uwatoko}
\affiliation{
The Institute for Solid State Physics, The University of Tokyo,
Chiba 277-8581, Japan
}

\author{Masashi~Tokunaga}
\affiliation{
The Institute for Solid State Physics, The University of Tokyo,
Chiba 277-8581, Japan
}

\date{\today}

\begin{abstract}
This study investigates the transport properties of a chiral elemental semiconductor tellurium (Te) under magnetic fields and pressure.
Application of hydrostatic pressure reduces the resistivity of Te, while its temperature dependence remains semiconducting up to 4 GPa, contrary to recent theoretical and experimental studies.
Application of higher pressure causes structural as well as semiconductor--metal transitions.
The resulting metallic phase above 4 GPa exhibits superconductivity at 2 K along with a noticeable linear magnetoresistance effect.
On the other hand, at ambient pressure, we identified metallic surface states on the as-cleaved (10$\bar{1}$0) surfaces of Te.
The nature of these metallic surface states has been systematically studied by analyzing quantum oscillations observed in high magnetic fields.
We clarify that a well-defined metallic surface state exists not only on chemically etched samples that were previously reported, but also on as-cleaved ones.
\end{abstract}

\maketitle

%\tableofcontents

\section{Introduction}

Strong spin-orbit interaction (SOI) realizes several types of spin-polarized electronic bands in materials not possessing spatial inversion symmetry.
The coupled spin and charge degrees of freedom in this class of materials result in non-trivial dynamics of spins and charges.
To elucidate the fundamentals of this coupling, it is crucial to conduct in-depth studies on simple materials;
in particular, materials having charge carriers with high mobility and low density provide a promising platform for the extraction of a non-trivial component.
In this context, elemental tellurium (Te) is a suitable material on which to carry out studies regarding anomalous phenomena.

\begin{figure}
\centering
\includegraphics[width=7.5cm]{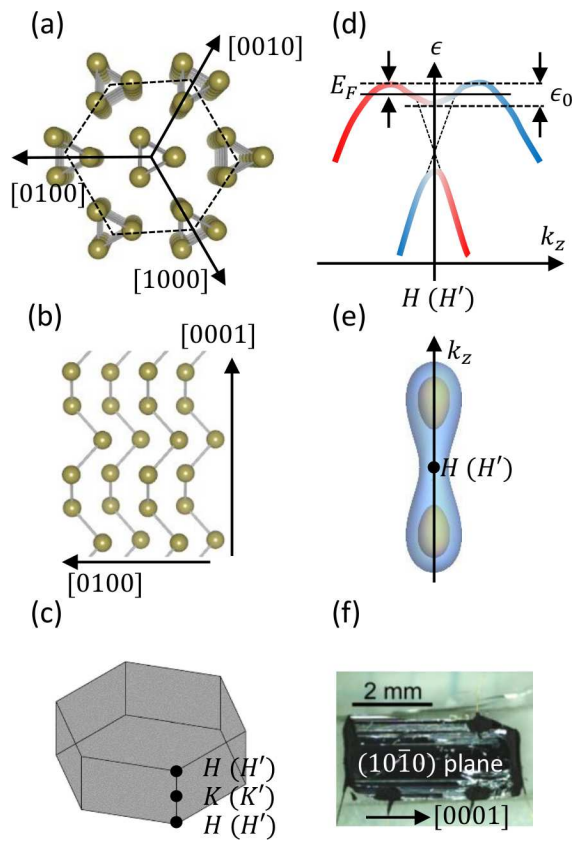}
\caption{
Crystal structure of Te viewed (a) from [0001] direction and (b) from [10$\bar{1}$0] direction.
(c) First Brillouin zone of Te.
(d) Schematic valence band structure of Te along the direction $k_z$ in the vicinity of the $H$ ($H'$) point with spin-orbit interaction.
Red and blue colors represent the spin polarization along the [0001] axis.
(e) Isoenergetic surfaces of Te at the $H$ ($H'$) point.
The inner and outer surfaces represent the cases of $E_F<\epsilon_0$ and $E_F>\epsilon_0$, respectively.
(f) Photograph of a Te sample used for transport properties measurements.
\label{fig_fig1}}
\end{figure}

At ambient pressure, Te is a narrow-gap semiconductor that forms a trigonal structure, as shown in Figs. \ref{fig_fig1}(a) and (b).
It consists of helices of covalently bonded Te atoms along the [0001] direction.
The helices are weakly bound by Van der Waals forces and form a chiral crystal structure with the space group $P3_121$ or $P3_221$
for the right-handed or left-handed helical axes, respectively.
The first Brillouin zone of Te is a hexagonal prism, as shown in Fig. \ref{fig_fig1}(c). The narrowest band gap of 0.32 eV \cite {Loferski} is located near the $H$ ($H'$) point.
Usually, Te shows $p$-type conduction without any doping, presumably due to holes provided by defects \cite{Takita}. 
The electronic structure of the valence band has been established both experimentally and theoretically.
According to the $\bm{k}\cdot \bm{p}$ theory \cite{Doi_I},
the energy dispersion of the valence band $E(\bm{k})$ is approximated by the following form:
\begin{equation}
E(\bm{k})=A (k_x^2+k_y^2)+B k_z^2 + \sqrt{S^2 k_z^2+4 \Delta_1^2},
\end{equation}
where $A=-3.4\times 10^{-15}$eV cm$^2$, $B=-4.6\times 10^{-15}$eV cm$^2$,
$S=2.67\times 10^{-8}$eV cm, and $\Delta_1=32.2$ meV \cite{Doi_II}.
As seen in Fig. \ref{fig_fig1}(d), $E(\bm{k})$ along the direction $k_z$ has the characteristic ``camel's back'' shape
with two maxima around the $H$ ($H'$) point.
Here, a strong SOI causes spin polarization, as indicated in red and blue in Fig. \ref{fig_fig1} (d), which represents parallel and antiparallel spins with respect to the [0001] direction, respectively.
Such a characteristic band structure, which reflects strong SOI, was directly observed by recent angle-resolved photoemission spectroscopy measurements \cite{Nakayama, Sakano}.
Within the $k_x$-$k_y$ plane, the spins are oriented radially from the $H$ ($H'$) point.
Such a spin texture suggests the emergence of magnetization induced by the charge current \cite{Shalygin}. Recent nuclear magnetic resonance experiments resolved the emergence of current-induced longitudinal magnetization, although its origin remains uncertain \cite{Furukawa}.
In the three-dimensional $k$-space, the isoenergetic surface of Te varies from a pair of spin-polarized ellipsoids in the case of $E_F<\epsilon_0$ [inner surfaces in Fig. \ref{fig_fig1}(e)] to "dumb-bell" in the case of $E_F>\epsilon_0$ [outer surface in Fig. \ref{fig_fig1}(e)], where $E_F$ represents the Fermi level, defined by the energy from the top of the valence band, and $\epsilon_0$ denotes the depth of the saddle, as shown in Fig. \ref{fig_fig1}(d).
The structure of the valence band has been investigated mainly by cyclotron/interband resonances and Shubnikov--de Haas (SdH) oscillation measurements in Sb-doped samples, which have carrier densities of 10$^{16}$-10$^{18}$ cm$^{-3}$ \cite{Radoff, Couder, Yoshizaki, Dubinskaya,Guthmann,Bresler_1969,Bresler_1970,Braun}. 

Application of hydrostatic pressure is useful for tuning of the band gap of Te. Te is known to show sequential structural phase transitions under pressure \cite{Akahama,Parthasarathy,Marini}.
The trigonal structure turns into the Te II structure at approximately 4 GPa. 
This phase is known to be metallic, and it demonstrates superconductivity at transition temperatures between 2 and 5 K \cite{Matthias}.
In addition, pressure-induced Weyl semimetal transition at $\sim 2.2$~GPa in the trigonal phase 
has been predicted theoretically \cite{Hirayama}, and recent experimental
data have been interpreted in favor of this scenario \cite{Ideue}.
Also, Agapito predicted that if we close the gap within the trigonal symmetry, Te becomes a strong topological insulator \cite{Agapito_2013}.
On the other hand, another high-pressure study using a diamond-anvil-cell reported that the band gap did not close up to 6 GPa \cite{Shchennikov}.
As mentioned above, it is still controversial whether the bandgap closes by pressure maintaining the trigonal structure.
Thus, high-pressure study with high hydrostaticity should be necessary to clarify this point.

Metallic states of pure Te can also be realized on the surface of Te at ambient pressure.
In 1971, von Klitzing and Landwehr observed SdH
oscillations in pure Te \cite{Klitzing}.
The observed metallic states were attributed to the surface accumulation layer created by the chemical etching process.
According to earlier SdH experiments on the etched (10$\bar{1}$0) surface of Te, the two 
observed SdH frequencies are attributed to the shallow and deep electronic subbands with $E_F\simeq \epsilon_0$ and $E_F\gg \epsilon_0$, respectively \cite{Berezovets_1991}. 

Here, we studied the magneto-transport properties of Te in high magnetic fields up to 55 T
at ambient pressure and high pressure up to 8 GPa at 0 T with cubic-anvil-type pressure cell.
In our measurements, pressure-induced metallization within the trigonal phase was not observed.
On the other hand, we identified SdH oscillations in the metallic surface state in Te $without$ chemical etching, which indicates the inevitable influence of this surface state when measuring the physical properties of Te in the atmosphere.

\section{Experiments}
Single crystals of Te investigated in this study were prepared by the Bridgman method.
Samples were prepared by cleaving  large ingots (typically 3 mm$\times$3 mm$\times$5 mm) in liquid nitrogen. 
For transport measurements, gold wires were attached to the cleaved shiny surfaces using carbon paste, as shown in Fig. \ref{fig_fig1}(f). 
Resistivity measurements under pressure up to 8 GPa were carried out mainly using a cubic-anvil-type pressure cell,
which can generate pressure with high hydrostaticity even after the solidification of the pressure media \cite{Mori}.
Glycerol were used as pressure media.
The magnetic field dependence of the resistivity at 5 GPa [Fig. \ref{fig_fig2}(d)] was performed with a combination of cubic-anvil-type pressure cell and superconducting magnet ($<$ 5 T, Cryomagnetics, Inc.)
Magnetoresistance measurements up to 14 T were carried out with the Physical Properties Measurement System
(PPMS, Quantum Design).
Magnetoresistance up to 55 T was measured using nondestructive pulse magnets (time duration of 36 ms) installed at The Institute for Solid State Physics, The University of Tokyo.
Field-angle dependence of magnetoresistance was studied using a sample rotator.
Transport measurements under pulsed magnetic fields were performed by a numerical lock-in technique
at a typical frequency of 100 kHz.

\section{Results and Discussion}

\begin{figure}
\centering
\includegraphics[width=7.5cm]{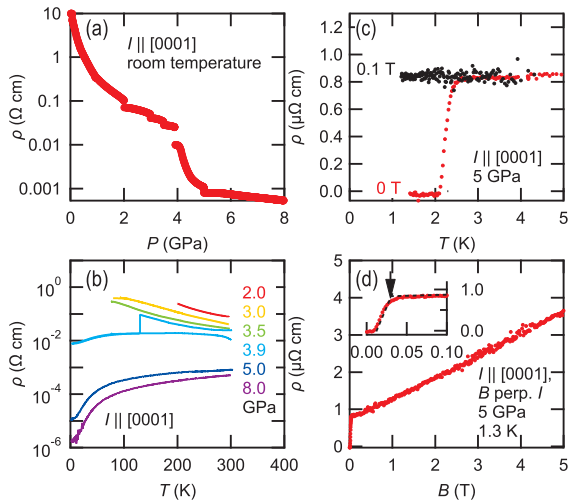}
\caption{
(a) Pressure dependence of the resistivity $\rho$ up to 8 GPa at room temperature.
(b) Temperature dependence of $\rho$ at various pressures.
(c) Temperature dependence of $\rho$ at 5 GPa in magnetic field of 0 and 0.1 T.
(d) Magnetic field dependence of $\rho$ up to 5 T at 5 GPa and 1.3 K.
The inset shows the magnified view of low-field region up to 0.1 T.
\label{fig_fig2}}
\end{figure}

Firstly, we focus on transport properties under pressure.
As mentioned previously, a recent theoretical calculation predicted the emergence of the Weyl semimetal phase
without inversion symmetry above 2.2 GPa.
Such metallization should be probed by the change in the temperature dependence of the resistivity
or the emergence of SdH oscillation under applied pressure.

\begin{figure}
\centering
\includegraphics[width=7.5cm]{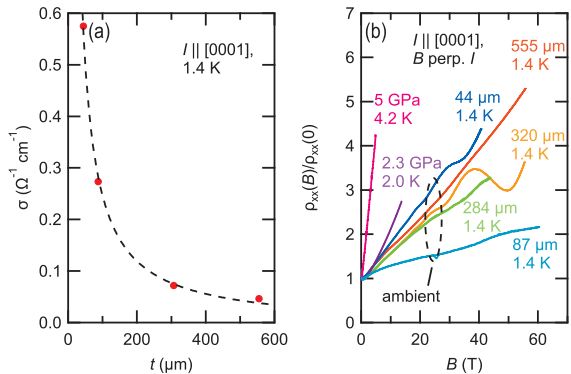}
\caption{
(a) Conductivity $\sigma=1/\rho$ of samples of different thickness $t$ at 1.4 K and zero magnetic field.
(b) Comparison of transverse magnetoresistance normalized by values in zero field.
\label{fig_fig5}}
\end{figure}

Figure \ref{fig_fig2}(a) shows the pressure dependence of the resistivity ($\rho$) at room temperature.
Abrupt decrease in resistivity was observed at approximately 4 GPa, which corresponds to the structural transition to the Te II phase as reported in previous studies \cite{Akahama,Parthasarathy, Hejny, Marini}.
Figure \ref{fig_fig2}(b) shows the temperature ($T$) dependence of resistivity at various pressures up to 8.0 GPa.
Although the resistivity decreases by more than two orders of magnitude at room temperature, we found that the $\rho$-$T$ curves  continued to depict semiconducting up to 3.9 GPa, contrary to the recent theoretical and experimental studies \cite{Hirayama, Ideue}.
At 2.3 GPa, we also measured the magnetoresistance at 2 K with $B \perp$ [0001] and $I \parallel$ [0001]
up to 14 T using piston-cylinder-type pressure cell and Daphne7373 oil as a pressure medium, which shows a monotonic increase in $\rho$ by a factor of 2.8 at 14 T without showing any signature of SdH oscillations as shown in Fig. \ref{fig_fig5}(b). 

The $\rho$-$T$ curves represented metallic character above 5.0 GPa. The ratio $\rho$(300 K)/$\rho$(2.4 K) increases to 315 at 8 GPa, indicating a highly conductive state.
The red curve in Fig. \ref{fig_fig2}(c) shows the temperature dependence of the $\rho$-$T$ curve at 5 GPa in a zero magnetic field. We observed a clear superconducting transition at a critical temperature of $\sim 2.3$ K. This superconducting state can be easily suppressed by application of a magnetic field ($B$) of 0.02 T, as shown in the inset of Fig. \ref{fig_fig2}(d). 
Some of the previous studies suggest that the Te II phase has a monoclinic structure without spatial inversion symmetry \cite{Akahama,Parthasarathy}.
We do not observe an enhancement of superconducting critical field, which is characteristic of non-centrosymmetric superconductors.
The normal state shows significant linear magnetoresistance, as shown in Fig. \ref{fig_fig2}(d), $\rho$(5 T)/$\rho$(0.05 T) = 4.4 at 1.3 K.

In classical electron-hole two-carrier model, the magnetoresistance is known to have quadratic magnetic field dependence in weak magnetic field and then saturate in sufficiently high magnetic field.
Large magnetoresistance is widely observed in compensated semimetals with high mobility carriers.
In elemental Te, each Te atom has an even number of electrons.
Therefore, the gapless state in the Te II phase should be a compensated semimetal.
This semimetallic state could be the essential point of the observed large positive magnetoresistance in this phase.
The linear magnetoresistance is expected to occur in the quantum limit state of a system having a linear energy dispersion \cite{Abrikosov} such as Dirac and Weyl semimetals.
However, the power of magnetoresistance can be affected by disorder effects \cite{Narayanan, Song}
and also the curvature of the relevant Fermi surface \cite{Awashima}.
We also note that the data shown in Fig. \ref{fig_fig2}(d) is not symmetrized as a function of $B$,
and hence, we cannot rule out possible contribution of the Hall resistance due to slight misalignment of the electrical contacts.
In order to discuss the origin for the linear magnetoresistance, we need additional information about the underlying electronic state.

\begin{figure}
\centering
\includegraphics[width=7.5cm]{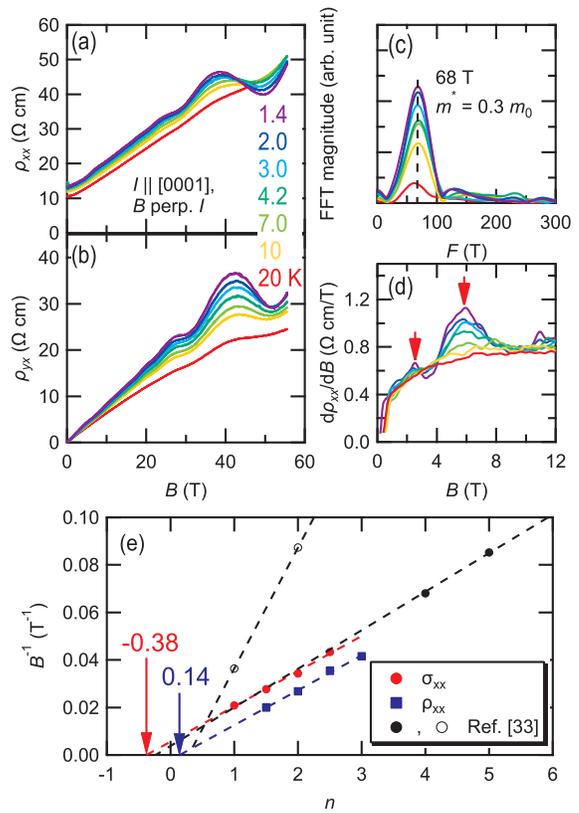}
\caption{
(a) Magnetoresistivity ($\rho_{xx}$) and (b) Hall resistivity ($\rho_{yx}$) up to 55 T
at several temperatures.
(c) FFT spectra of oscillations superimposed on $\rho_{xx}$.
(d) Magnified view of $d\rho_{xx}/dB$ below 12 T.
The red arrows indicate the SdH oscillation, discernible only below 6 T.
(e) Landau level fan diagram constructed from $\sigma_{xx}$ (red) and $\rho_{xx}$ (blue).
The peaks are plotted for integer $n$.
Two SdH components mentioned in the previous data of Berezovets \textit{et al.} \cite{Berezovets_1995} are also shown
by closed and open black circles, which are constructed from peaks in $-d^2\rho_{xx}/dB^2$.
\label{fig_fig3}}
\end{figure}

As shown above, we could not realize bulk metallic states in trigonal Te under pressure.
In the case of black phosphorus, which also has a direct band gap of $\sim$0.3 eV at ambient pressure,
absorption of alkali-ions can close the gap at the surface instead of the application of hydrostatic pressure \cite{Kim}.
Also in Te, the emergence of a metallic surface state is reported for the chemically etched crystal.
We therefore focus on the transport properties in the surface state in the following.

We measured the magnetoresistivity ($\rho_{xx}$) and the Hall resistivity ($\rho_{yx}$) in pulsed magnetic fields up to 55 T at various $T$ from 1.4 to 20 K. Here, magnetic fields were applied normal to the (10$\bar{1}$0) plane, while the currents flowed along the [0001] axis. As shown in Figs. \ref{fig_fig3} (a) and (b), the modulations superimposed on both $\rho_{xx}$ and $\rho_{yx}$ can be recognized above 20 T and are systematically damped with increasing temperature.
We also confirmed that this modulation is periodic as a function of $B^{-1}$. Thus, we attribute this structure to the SdH oscillation.
Figure \ref{fig_fig3} (c) shows the fast Fourier transformation (FFT) spectra of oscillations on $\rho_{xx}$. A single peak was detected at $F=68$ T, and a light cyclotron mass of $m^*=0.3$ $m_0$,
where $m_0$ represents the free electron mass, is estimated from the temperature dependence of the spectra.

Here, we consider dimensionality of the Fermi surface. If the surface is three-dimensional, 
the carrier density ($n_{3D}$) is related to the SdH frequency $F$ as
\begin{equation}
n_{3D}=\dfrac{g_s g_v}{6\pi^2}\left( \dfrac{2eF}{\hbar}\right)^{3/2},
\label{eq_n3d}
\end{equation}
where $e$ and $\hbar$ represent the elemental charge and reduced Planck constant,
respectively.
$g_s$ and  $g_v$ represent the spin- and valley-degeneracy, respectively.
$g_s=1$, since the top of the valence band has no spin-degeneracy, as mentioned previously.
$g_v$ is equal to either 2 or 4 when the Fermi surface is a dumb-bell or a pair of ellipsoids, respectively.
In both cases, however, by substituting $F=68$ T into Eq. (\ref{eq_n3d}),
we found that $n_{3D}$ is of the order 10$^{17}$ cm$^{-3}$.
This carrier density is considerably larger than that estimated from the slope of $\rho_{yx}$ in a weak magnetic field (typically $10^{15}$ cm$^{-3}$)
and that from previous reports ($1.9-6.6 \times 10^{14}$ cm$^{-3}$) \cite{Takita}.
In the case of a two-dimensional metal, the carrier density ($n_{2D}$) is represented as
\begin{equation}
n_{2D}=g_s g_v\dfrac{eF}{2\pi  \hbar}.
\end{equation}
We can estimate $n_{2D}$ as $3.3\times 10^{12}$ cm$^{-2}$ ($6.6\times 10^{12}$ cm$^{-2}$)
with $g_s g_v=2$ ($g_s g_v=4$), which is the same order of magnitude as for surface states in the chemically etched Te \cite{Berezovets_1991}.
The above estimation indicates the existence of a well-defined two-dimensional Fermi surface in our as-cleaved sample
without usage of chemical etching.
 
In addition to this predominant SdH oscillation above 20 T, small hump-like structures in $\rho_{xx}$ are discernible at low-field, which are clearly visible in $d\rho_{xx}/dB$ as indicated by red arrows in Fig. \ref{fig_fig3}(d). 
These structures can be attributed to another oscillation component of the smaller Fermi surface.
From the interval of two peaks and temperature dependence of the peak height at 6 T, the frequency and the cyclotron mass are estimated to 4.6 T and 0.06$m_0$, respectively.
The observed two oscillation components with $F=68$ T and 4.6 T are comparable with the SdH oscillations regarded to stem from
deep and shallow subbands in chemically etched Te \cite{Berezovets_1991}.

From the present data, we constructed the Landau level fan diagram,
as shown in Fig. \ref{fig_fig3}(e).
Here the integer Landau indices are assigned to the peaks of $\sigma_{xx} (=\frac{\rho_{xx}}{\rho_{xx}^2+\rho_{yx}^2})$. The black symbols represent the data quoted from the literature,
in which the peaks in $-d^2\rho_{xx}/dB^2$ are assigned as the level crossing fields \cite{Berezovets_1995}.
The SdH oscillation with $F=68$ T shows a horizontal intercept of $\sim-0.38$, which is similar to that of the deep subband discussed in Ref. \cite{Berezovets_1995}.
We cannot evaluate the phase shift for SdH with $F=4.6$ T since the oscillation is quite weak.

\begin{figure}
\centering
\includegraphics[width=7.5cm]{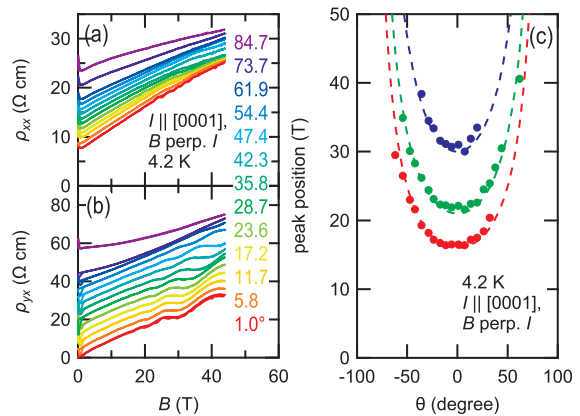}
\caption{
(a) $\rho_{xx}$ and (b) $\rho_{yx}$ in different magnetic field directions
with angles between the magnetic field and the normal vector of (10$\bar{1}$0) surface.
The field direction is always normal to the electric current.
The data are offset vertically for clarity.
(c) Angular dependence of the peak positions in $d\rho_{yx}/dB$.
The dotted line represents the curve $B_p=A/\cos \theta$ for $A=16$ (red), 21 (green), and 30 (blue), respectively.
\label{fig_fig4}}
\end{figure}

Next, we show the field angular dependence of the SdH oscillations in pulsed magnetic fields.
Figures \ref{fig_fig4}(a) and (b) represent $\rho_{xx}$ and $\rho_{yx}$ at different magnetic field directions.
Here electric currents are applied along the [0001] axis,
and magnetic fields are always applied normal to the current direction.
$\theta$ is defined as the angle between the magnetic field and the normal vector of the cleaved (10$\bar{1}$0) plane.
In this sample, SdH oscillations appear more clearly in $\rho_{yx}$ than in $\rho_{xx}$.
Figure \ref{fig_fig4}(c) shows the angular dependence of the SdH peaks in $d\rho_{yx}/dB$ at 4.2 K.
If only the top sample surface is related to the observed SdH oscillations, 
peak positions ($B_p$) should be scaled by the magnetic field component normal to the surface; thus, an angular dependence $B_p \propto 1/\cos \theta$ is obtained.
As shown in Fig. \ref{fig_fig4}(c), the angular dependence of the peak positions is reasonably reproduced by this formula, which confirms that the SdH oscillation
originates from the two-dimensional metallic layer on the sample surface.

The dependence of resistivity on thickness further confirms the contribution of metallic surface conductivity
in pristine Te.
Figure \ref{fig_fig5}(a) shows the conductivity $\sigma=1/\rho$ of as-cleaved samples with different thicknesses at 1.4 K and zero magnetic field.
In conventional bulk materials, the conductivity is constant regardless of the sample dimensions.
However, in the present case, it clearly increases as the thickness decreases.
Here, we assume that both the semiconducting bulk and the metallic surface contribute to $\sigma$.
Then, $\sigma$ is inversely proportional to the thickness: $\sigma=\sigma_b+G/t$, where $\sigma_b$, $G$, and $t$ represent
the bulk conductivity, surface sheet conductivity, and thickness of the sample, respectively.
Our data are reasonably reproduced by the above model with
$G=0.00254\pm0.00006$ $\Omega^{-1}$ at 1.4 K and
$\sigma_b=–0.008\pm0.008$ $\Omega^{-1}$ cm$^{-1}$,
as shown by the dotted line in Fig. \ref{fig_fig5}.
This systematic change supports the parallel conduction of semiconducting bulk and metallic surface channels
in as-cleaved Te samples.

Finally, we summarize the magnetoresistance effect in our study.
Figure \ref{fig_fig5}(b) shows the transverse magnetoresistance of several samples at ambient and hydrostatic pressure.
Here, the vertical axis is normalized by resistivity values in a zero magnetic field.
The magnetoresistance effect changes less systematically against the sample thickness at ambient pressure
and is slightly enhanced by applying a pressure of 2.3 GPa.
In the Te II phase, we can recognize a significant enhancement of magnetoresistance,
which may imply the existence of high-mobility carriers in the compensated semimetallic phase.

Our results indicate that a well-defined metallic surface state exists on samples without a chemical etching process.
The underlying local band structure should be determined by future spectroscopic measurements.
Owing to its high mobility, the contribution of the surface state is not negligible in bulk measurements, which is evident from the observation
of SdH oscillation at a temperature of a few Kelvin and the thickness-dependent conductivity.
This factor should be always considered regarding any measurement of transport properties for bulk Te.

\section{Conclusions}
We investigated the transport properties of elemental tellurium.
Under applied hydrostatic pressure up to 8 GPa, we observed a structural phase transition from the trigonal to the Te II structure at 4 GPa with a sharp drop in resistivity at room temperature.
The semiconducting temperature dependence of resistivity is retained up to the structural phase transition,
whereas metallic character is seen in the Te II phase.
We cannot find any indications of a semiconductor-metal transition within the trigonal structure reported by recent theoretical and experimental studies.
At ambient pressure, we observed clear Shubnikov--de Haas oscillations on the as-cleaved (10$\bar{1}$0) surface in pulsed magnetic fields up to 55 T.
The oscillation contains two frequencies (68 T and 4.6 T).
These results indicate that the metallic surface state makes a significant contribution to the studies of transport properties in bulk crystals of Te, even on as-cleaved surfaces without chemical etching.

\begin{acknowledgments}
We thank K. Behnia for helpful comments on the electronic state in the Te II phase.
This study was supported by JSPS KAKENHI Grant Numbers 19K14660 and 19H01850.
\end{acknowledgments}

\bibliography{reference}% Produces the bibliography via BibTeX.

\end{document}